# MULTISCALING IN EXPANSION-MODIFICATION SYSTEMS: AN EXPLANATION FOR LONG RANGE CORRELATION IN DNA.

**Ricardo Mansilla**[*,**]
**(mansilla@scifunamifisicacu.unam.mx)**
**Germinal Cocho**[**]
**(cocho@fenix.ifisicacu.unam.mx)**

*** Faculty of Mathematics and Computer Science, University of Havana, Cuba.**
**** Institute of Physics, UNAM, Mexico.**

Abstract.

Based on a model first studied in [Li, 1991a], properties of correlation function for expansion-modification systems are developed. The existence of several characteristic exponents is proved. The relationship of this fact with long-range correlation in DNA is stablished. Comparison between theoretical exponents and those obtained from simulation and real sequences are also showed.

Whom correspondence should be sent: mansilla@scifunam.ifisicacu.unam.mx.

## 1 Introduction.

The discovery of DNA molecule has revolutionised our way of thinking about biological evolution [10]. One of the most important challenges in our time is the understanding of its functioning, because, notwithstanding the huge amount of sequenced base pairs, the rate of interpretation of this data is lagging behind the rate of acquisition.

One of the most fruitful lines of research in recent years is related to long-range correlation in DNA [Borsnitk, 1993], [Buldyrev, 1995], [Burks & Farmer, 1984], [Chatzidimitrou-Dreismann & Larhamar, 1993], [Karlin & Brendel, 1993], [Larhamar & Chatzidimitrou-Dresimann, 1993], [Li, 1987], [Li, 1989], [Li, 1992], [Li & Kaneko, 1992], [Li & Kaneko, 1992a], [Mansilla & Mateo-Reig, 1995], [Miramontes, 1992], [Nee, 1992], [Peng et el., 1992], [Peng et al., 1993], [Prabhu & Claverie, 1992], [Voss, 1992]. Most of these papers report experimental evidences on long-range correlation [Borsnitk, 1993], [Buldyrev, 1995], [Chatzidimitrou-Dreismann & Larhamar, 1993], [Karlin & Brendel, 1993], [Li, 1992], [Li & Kaneko, 1992], [Peng et el., 1992], [Prabhu & Claverie, 1992] and just a few develop theoretical models addressed to explaining the above mentioned property [Li, 1991], [Li, 1992], [Mansilla & Mateo-Reig, 1995]. In [Li, 1989], a model is exposed further studied in [Li, 1991a], which grasps the main features in the formation of long-range correlation in DNA: point mutation and insertion (see, for example, [Mansilla & Mateo-Reig, 1995]).

In [Li, 1991a] a behaviour of the correlation function of the form $1/d^c$ is assumed , where $d$ is the distance between symbols in the sequence and $c$ is supposed to depend on the probability $p$ of mutation. In [Chatzidimitrou-Dreismann & Larhamar, 1993] a behaviour of the form $d^{\varphi(d)}$ is proposed for a magnitude inversely proportional to correlation function. In [Buldyrev et al., 1995] and [Li & Kaneko, 1992] the correlation function for a group of sequences of GENBANK is calculated and the behaviour of these functions suggests the existence of more than one exponent, e.g. a behaviour of the form $\sum_i 1/d^{c_i}$ .

The results of simulations in [Li, 1989] also suggest the above mentioned behaviour.

The aim of this paper is to prove that for the model proposed in [Li, 1991a], the correlation function has a behaviour of the form: $\sum_i K_i / d^{\varphi_i(d)}$. We obtain the constants $K_i$ and asymptotic upper and lower bounds for the functions $\varphi_i(d)$. We also prove that one of these exponents has a more important contribution to correlation functions than the others and show its fitness with respect to those obtained in simulations.

The structure of this paper is as follow: In Sec. 1 some preliminary definitions are reviewed. In Sec. 2 the fundamental mathematical results are developed. Among them the most important are Eq. (2.6) and Eq. (2.11). In Sec. 3 upper and lower bounds for the eigenvalues of the matrix $M(i_{k-1}, i_k)$ which appear in Eq. (2.6) are obtained. As stated in (3.10a) and (3.10b) those bounds depend on the sum $S(d,n)$ and the product $M(d,n)$ of elements of some multiindex. In Sec. 4 upper and lower bounds for $S(d,n)$ and $M(d,n)$ are obtained. In Sec. 5, for $\delta_j(i_1,\ldots,i_n)$ defined in (2.12), uniform upper and lower bounds on sets $G(d,n)$ are obtained. From this result, asymptotic expressions are obtained for the exponent in correlation function. In Sec. 6 the exponent of the major contributing term is calculated, and this theoretical exponent and that obtained from simulations are compared in a plot. In Sec. 7 the results are discussed and Sec. 8 is for conclusion.

## 1 Some Preliminary Definitions.

The evolution of prebiotic nucleotide sequences in an instance in which two competing processes play an important role in determining the statistical properties of the sequences.

Among the group of modifications that DNA sequences suffer, the replications and point mutations are, in some sense, antagonistic. Replications add strings of chain in other sites creating long range correlation, while point mutations tend to destroy them. If the prebiotic evolution contained only replications, the limiting sequence would be periodic; if the point mutation rate was too high, the limiting sequence would be random. Only when the two processes are in an appropiate balance, can the nucleotide sequences show nontrivial long range correlation as observed in Nature.

In [Li, 1991a] a model is proposed which grasps the main features of this processes. Let $x^t = \cdots \alpha_0^t \alpha_1^t \cdots$ be a binary sequence. It is mapped in $x^{t+1} = \cdots \alpha_0^{t+1} \alpha_1^{t+1} \cdots$ using the following rules: each symbol $\alpha_i^t$ changes to two identical symbols with probability $1-p$ and switches to the other symbol with probability $p$, i.e.,

$$0 \rightarrow \begin{cases} 00 & 1-p \\ 1 & p \end{cases} \qquad ; \qquad 1 \rightarrow \begin{cases} 11 & 1-p \\ 0 & p \end{cases}$$

A particular realisation of this rewriting process is shown in Fig. 1 of [Li, 1991a].

We borrow some definitions. Let $P_{\alpha,\beta}^t(d)$ be the joint probability for having the symbol pair $\alpha$ , $\beta$ separated by the distance $d$ .

Assuming that the transition probability from an $\alpha'$ , $\beta'$ pair initially at distance $d'$ to an $\alpha$ , $\beta$ pair at distance $d$ is $T(\alpha', \beta', d' \rightarrow \alpha, \beta, d)$, the joint probabilities satisfy the following dynamical equation:

$$\begin{bmatrix} P_{0,0}^{t+1}(d) \\ P_{0,1}^{t+1}(d) \\ P_{1,0}^{t+1}(d) \\ P_{1,1}^{t+1}(d) \end{bmatrix} = \sum_{d'=[d/2]}^{d} \begin{bmatrix} T(00 \to 00) & . & . & T(11 \to 00) \\ . & . & . & . \\ . & . & . & . \\ T(00 \to 11) & . & . & T(11 \to 11) \end{bmatrix} \begin{bmatrix} P_{0,0}^{t}(d') \\ P_{0,1}^{t}(d') \\ P_{1,0}^{t}(d') \\ P_{1,1}^{t}(d') \end{bmatrix} \qquad (1.1)$$

here we have written $T(\alpha',\beta' \to \alpha,\beta)$ instead of $T(\alpha',\beta',d' \to \alpha,\beta,d)$ for simplicity. From now on, the square brackets [.] stand for integer part of number.

The transition probabilities $T(\alpha',\beta',d' \to \alpha,\beta,d)$ can be grouped into three types:

$T_0(d',d,p)$: Keep both symbols unchanged, for instance: $T(0,0 \to 0,0)$.

$T_1(d',d,p)$ : Change one symbols, for instance: $T(0,1 \to 1,1)$.

$T_2(d',d,p)$ : Change both symbols, for instance: $T(1,1 \to 0,0)$.

Hence the Eq. (1.1) can be written:

$$P^{t+1}(d) = \sum_{d'=[d/2]}^{d} T(d',d,p)P^{t}(d') \qquad (1.2)$$

where:

$$P^{t}(d) = \begin{bmatrix} P_{0,0}^{t}(d) \\ P_{0,1}^{t}(d) \\ P_{1,0}^{t}(d) \\ P_{1,1}^{t}(d) \end{bmatrix}$$

$$T(d',d,p) = \begin{bmatrix} T_0 & T_1 & T_1 & T_2 \\ T_1 & T_0 & T_2 & T_1 \\ T_1 & T_2 & T_0 & T_1 \\ T_2 & T_1 & T_1 & T_0 \end{bmatrix}$$

In the above matrix we have written $T_s$ instead of $T_s(d',d,p)$ for simplicity.

Suppose there is a time invariant condition in the $t \to +\infty$ limit and the superscript can be dropped. Then we can write Eq. (1.2) in the following form:

$$P(d) = \sum_{d'=[d/2]}^{d} T(d', d, p) P(d') \tag{1.3}$$

or:

$$P(d) = (I - T(d,d))^{-1} \sum_{d'=[d/2]}^{d-1} T(d', d, p) P(d') \tag{1.4}$$

In Sec. 2, starting from Eq. (1.4) we obtain a closed expression for $P(d)$.

## 2 Some Fundamental Mathematical Results.

**Definition 2.1**: Let $d \ge n > 1$ be integers. Let $G(d,n)$ be the set of elements of the form $(i_1, \ldots, i_n) \in \{1, \ldots, d\}^n$ which hold the following conditions:

a) $i_1 = 1$, $i_n = d$.

b) $i_1 < i_2 < \ldots < i_n$.

c) For every $l = 1, \ldots, n-1$ : $[i_{l+1}/2] \le i_l$.

Examples of such sets are:

$G(3,2) = \{(1,3)\}$
$G(4,3) = \{(1,2,4), (1,3,4)\}$
$G(d,d) = \{(1,2,\ldots,d)\}$
$G(4,2) = \varnothing$

In general: $G(d,2)=\varnothing$ for every $d \geq 4$. To prove this, let us note that $G(d,2)=\{(1,d)\}$ and as the condition c) of Definition 2.1 must be hold, then $[d/2] \leq 1$ , which is only possible for $d \leq 3$.

**Definition 2.2**: Let $d > k \geq n > 1$ be integers which satisfy the following conditions:

a) $[d/2] \leq k$.

b) $n \geq [\log_2 k]+1$.

Let us denote $(k,d) \wedge G(k,n)$ the set of elements of the form $(i_1,\ldots,i_n,d)$ such that: $(i_1,\ldots,i_n) \in G(k,n)$.

**Lemma 2.3**: Let $d > n > 1$. Denote:

$u(d,n) = \min\{d-1, 2^n - 1\}$
$l(d,n) = \max\{[d/2], n\}$

Then:

$$G(d,n+1) = \bigcup_{l(d,n)}^{u(d,n)} (k,d) \wedge G(k,n) \qquad (2.1)$$

**Proof**: Consider $(i_1,\ldots,i_n,d) \in (k,d) \wedge G(k,n)$. It implies that $k \geq n$ from the Definition 2.1; $[d/2] \leq i_n = k$ from condition a) of Definition 2.2; $k \leq d-1$ from Definition 2.2 and $k \leq 2^n - 1$ from condition b) of Definition 2.2. Hence we have $l(d,n) \leq k \leq u(d,n)$. Besides $(i_1,\ldots,i_n) \in G(k,n)$. Let us show that $(i_1,\ldots,i_n,d) \in G(d,n+1)$. Obviously $(i_1,\ldots,i_n,d) \in \{1,\ldots,d\}^{n+1}$; $i_1 = 1$ because $(i_1,\ldots,i_n) \in G(k,n)$ and also $i_{n+1} = d$. The above guarantees condition a) of Definition 2.1. Besides $i_1 <\ldots< i_n$ because $(i_1,\ldots,i_n) \in G(k,n)$ and $k = i_n < d$. Hence, condition b) of Definition 2.1 also holds. Lastly, condition c) of Definition 1 holds for $l = 1,\ldots,n-1$ because $(i_1,\ldots,i_n) \in G(k,n)$.

From condition a) of Definition 2.2, condition c) of Definition 2.1 holds for $l = n$. Hence $(i_1,\ldots,i_n,d) \in G(d,n+1)$. Let us see the opposite inclusion.

Let $(i_1,\ldots,i_{n+1}) \in G(d,n+1)$. It implies that $i_{n+1} = d$. Let $k$ be the element $i_n$. From condition c) of Definition 2.1 we have $[d/2] \le k$, therefore, condition a) of Definition 2.2 holds. Obviously $k \le d-1$. On the other hand, $k \ge n$ because $1 = i_1 <\ldots< i_n = k$. Let us prove that $(i_1,\ldots,i_n) \in G(k,n)$. First, $(i_1,\ldots,i_n) \in \{1,\ldots,k\}^n$ because $(i_1,\ldots,i_{n+1}) \in G(d,n+1)$. Consequently condition a) of Definition 2.1 holds. As we have said condition b) also holds. The condition c) of Definition 2.1 is true for $l = 1,\ldots,n-1$ because once again $(i_1,\ldots,i_{n+1}) \in G(d,n+1)$. All the above implies that $(i_1,\ldots,i_n) \in G(k,n)$ and therefore $n \ge [\log_2 k]+1$. Hence $l(d,n) \le k \le u(d,n)$.

As $(i_1,\ldots,i_n) \in G(k,n)$ then $(i_1,\ldots,i_{n+1}) \in (k,d) \wedge G(k,n)$ for certain $k$. But this implies that:

$$G(d,n+1) \subset \bigcup_{l(d,n)}^{u(d,n)} (k,d) \wedge G(k,n)$$

This completes the proof.

**Remarks**:

1) If $u(d,n) = 2^n - 1$, then for $2^n - 1 < k \le d-1$ we have $G(k,n) = \varnothing$. It is not difficult to prove that $G(k,n) \ne \varnothing$ if and only if $n \ge [\log_2 k]+1$. Hence, if $2^n - 1 < k$ we have $G(k,n) = \varnothing$.

2) Let us note that it is not possible $u(d,n) = 2^n - 1$ and $l(d,n) = n$. If $d-1 > 2^n - 1$ then $[d/2] > 2^{n-1}$, but $2^{n-1} \geq n$ for $n \geq 2$, therefore it is impossible that $[d/2] < n$.

From the above remarks, we obtain that the following equation:

$$G(d,n+1) = \bigcup_{k=n}^{d-1} (k,d) \wedge G(k,n) \qquad (2.2)$$

is also true.

**Corollary 2.4**: Let: $\theta(d,n) = cardG(d,n)$. Then we have:

$$\theta(d,n+1) = \sum_{k=l(d,n)}^{u(d,n)} \theta(k,n) \qquad (2.3)$$

**Proof**: It is straightforward from Eq. (2.1) and the fact that sets $G(d,n)$ do not intersect each other.

**Remark**: From the remarks below Lemma 2.3, the following expression:

$$\theta(d,n+1) = \sum_{k=n}^{d-1} \theta(k,n) \qquad (2.4)$$

is also true.

## TABLE I
## Values of cardinals from sets G(d,n)

| d\n | 2 | 3 | 4 | 5 | 6 | 7 | 8 | 9 | 10 | 11 | 12 | 13 | 14 |
|---|---|---|---|---|---|---|---|---|---|---|---|---|---|
| 2 | 1 | | | | | | | | | | | | |
| 3 | 1 | 1 | | | | | | | | | | | |
| 4 | 0 | 2 | 1 | | | | | | | | | | |
| 5 | 0 | 2 | 3 | 1 | | | | | | | | | |
| 6 | 0 | 1 | 5 | 4 | 1 | | | | | | | | |
| 7 | 0 | 1 | 6 | 9 | 5 | 1 | | | | | | | |
| 8 | 0 | 0 | 6 | 15 | 14 | 6 | 1 | | | | | | |
| 9 | 0 | 0 | 6 | 21 | 29 | 20 | 7 | 1 | | | | | |
| 10 | 0 | 0 | 4 | 26 | 50 | 49 | 27 | 8 | 1 | | | | |
| 11 | 0 | 0 | 4 | 30 | 76 | 99 | 76 | 35 | 9 | 1 | | | |
| 12 | 0 | 0 | 2 | 31 | 105 | 175 | 175 | 111 | 44 | 10 | 1 | | |
| 13 | 0 | 0 | 2 | 33 | 136 | 280 | 350 | 286 | 155 | 54 | 11 | 1 | |
| 14 | 0 | 0 | 1 | 30 | 165 | 415 | 630 | 636 | 441 | 209 | 65 | 12 | 1 |

Table I show the values of $\theta(d,n)$ for $2 \le d \le 13$ and $2 \le n \le 13$. It can be proved (see Appendix A for the details) that:

$$\theta(d,k) \le \frac{(2d-k-3)}{k-3}\binom{d-4}{d-k} \qquad (2.5)$$

Let us denote by: $I(d) = (I - T(d^{'},d))^{-1}$ and $M(d^{'},d) = I(d)T(d^{'},d)$. Then we have the following:

**Theorem 2.10**: For every $d \in \mathrm{N}$ , $d \ge 2$:

$$P(d) = \left\{ \sum_{n=2}^{d} \sum\nolimits_{(i_1,\ldots,i_n)\in G(d,n)} M(i_{n-1},i_n)\ldots M(i_1,i_2) \right\} P(1) \qquad (2.6)$$

**Proof**: It will be by induction on $d$. The property is true for $d = 3$:

$$P(3) = \left\{ \sum\nolimits_{(i_1,i_2)\in G(3,2)} M(i_1,i_2) + \sum\nolimits_{(i_1,i_2,i_3)\in G(3,3)} M(i_2,i_3)M(i_1,i_2) \right\} P(1)$$

Let suppose that it is true for $k = 2,\ldots,d-1$ and prove that it is also true for $k = d$.

Then:

$$P(d) = I(d)\sum_{k=2}^{d-1} T(k,d)P(k)$$

$$= I(d)\sum_{k=2}^{d-1} T(k,d)\left\{\sum_{n=2}^{k}\sum\nolimits_{(i_1,\ldots,i_n)\in G(k,n)} M(i_{n-1},i_n)\ldots M(i_1,i_2)\right\}P(1)$$

$$= \left\{\sum_{k=2}^{d-1}\sum\nolimits_{n=2}^{k}\sum\nolimits_{(i_1,\ldots,i_n)\in Gk,n)} M(k,d)M(i_{n-1},i_n)\ldots M(i_1,i_2)\right\}P(1) \quad (2.7)$$

Let us denote by:

$$\sigma(k,n) = \sum\nolimits_{(i_1,\ldots,i_n)\in G(k,n)} M(k,d)M(i_{n-1},i_n)\ldots M(i_1,i_2)$$

Then it is not difficult to see that:

$$\sum_{k=2}^{d-1}\sum\nolimits_{n=2}^{k}\sigma(k,n) = \sum_{n=2}^{d-1}\sum\nolimits_{k=n}^{d-1}\sigma(k,n)$$

Hence Eq. (2.7) can be written as:

$$P(d) = \left\{\sum_{n=2}^{d-1}\sum\nolimits_{k=n}^{d-1}\sum\nolimits_{(i_1,\ldots,i_n)\in G(k,n)} M(k,d)M(i_{n-1},i_n)\ldots M(i_1,i_2)\right\}P(1) \quad (2.8)$$

Now the multindex of the product $M(k,d)M(i_{n-1},i_n)\ldots M(i_1,i_2)$ is $(i_1,\ldots i_n,d)$. From Lemma 2.3 we have: $(i_1,\ldots i_n,d)\in(k,d)\wedge G(k,n)\subset G(d,n+1)$. From Eq. (2.2) we can write:

$$\sum_{(i_1,\ldots i_{n+1})\in G(d,n+1)} M(i_n,i_{n+1})\ldots M(i_1,i_2) = \sum_{k=n}^{d-1}\sum_{(i_1,\ldots i_n)\in G(k,n)} M(k,n)M(i_{n-1},i_n)\ldots M(i_1,i_2)$$

Therefore, the Eq. (2.8) can be written:

$$P(d) = \left\{\sum_{n=2}^{d-1}\sum\nolimits_{(i_1,\ldots i_{n+1})\in G(d,n+1)} M(i_n,i_{n+1})\ldots M(i_1,i_2)\right\}P(1)$$

Now making the change of variable $n = m-1$ the above expression can be written as:

$$P(d) = \left\{\sum_{m=3}^{d}\sum\nolimits_{(i_1,\ldots i_m)\in G(d,m)} M(i_{m-1},i_m)\ldots M(i_1,i_2)\right\}P(1)$$

We could add the term for $m=2$ because $G(d,2)=\varnothing$ for every $d \geq 4$ and the inner sum would be zero. Finally:

$$P(d)=\left\{\sum_{n=2}^{d}\sum\nolimits_{(i_1,\ldots i_n)\in G(d,n)} M(i_{n-1},i_n)\ldots M(i_1,i_2)\right\}P(1)$$

where we have changed $m$ by $n$. This completes the proof.

**Remarks**:

1) Some terms of Eq. (2.6) are equal to zero. As we have pointed out $G(d,n)\neq\varnothing$ if and only if $n \geq [\log_2 d]+1$. Besides:

$\sum\limits_{(i_1,\ldots i_n)\in G(d,n)} M(i_{n-1},i_n)\ldots M(i_1,i_2)=0$ if $n<[\log_2 d]+1$ because in each product $M(i_{n-1},i_n)\ldots M(i_1,i_2)$ there exists at least a couple $i_{r-1},i_r$ which does not satisfy condition c) of Definition 2.1 and therefore $M(i_{r-1},i_r)=0$. This explains why we could add the term for $n=2$ at the end of the proof of Theorem 2.10. Hence the following expression remains true:

$$P(d)=\left\{\sum_{n=[\log_2 d]+1}^{d}\sum\nolimits_{(i_1,\ldots i_n)\in G(d,n)} M(i_{n-1},i_n)\ldots M(i_1,i_2)\right\}P(1) \quad (2.9)$$

2) Note that $P(d)$ depends on $P(1)$. In sequences of four symbols (as DNA sequences), $P(1)$ is related to the dimers structure. As we have shown in [Mansilla et al., 1993] and [Mansilla & Mateo-Reig, 1995] $P(1)$ distinguishes the non coding regions of DNA molecule. That is why we use it in [ Mansilla & Mateo-Reig, 1995] as fitness function for an evolutionary model studied there.

Our next step is to study the structure of matrix $T(d',d,p)$. It can be proved that:

$$T(d',d,p)=\begin{bmatrix} T_0 & T_1 & T_1 & T_2 \\ T_1 & T_0 & T_2 & T_1 \\ T_1 & T_2 & T_0 & T_1 \\ T_2 & T_1 & T_1 & T_0 \end{bmatrix}=Q\begin{bmatrix} \pi_1 & 0 & 0 & 0 \\ 0 & \pi_2 & 0 & 0 \\ 0 & 0 & \pi_2 & 0 \\ 0 & 0 & 0 & \pi_3 \end{bmatrix}Q^{-1}=QDQ^{-1}$$

where:

$$\pi_1=T_0+2T_1+T_2 \quad ; \quad \pi_2=T_0-T_2 \quad ; \quad \pi_3=T_0-2T_1+T_2 \qquad (2.10)$$

$$Q=\begin{bmatrix}1 & -1 & 0 & 1\\ 1 & 0 & -1 & -1\\ 1 & 0 & 1 & -1\\ 1 & 1 & 0 & 1\end{bmatrix} \quad ; \quad Q^{-1}=\begin{bmatrix}1/4 & 1/4 & 1/4 & 1/4\\ -1/2 & 0 & 0 & 1/2\\ 0 & -1/2 & 1/2 & 0\\ 1/4 & -1/4 & -1/4 & 1/4\end{bmatrix}$$

In the same way:

$$I(d)=Q\begin{bmatrix}(1-v_1)^{-1} & 0 & 0 & 0\\ 0 & (1-v_2)^{-1} & 0 & 0\\ 0 & 0 & (1-v_2)^{-1} & 0\\ 0 & 0 & 0 & (1-v_3)^{-1}\end{bmatrix}Q^{-1}=Q\Xi Q^{-1}$$

where $v_1, v_2, v_3$ are the eigenvalues of the matrix $T(d,d,p)$. Using the above expression, Eq. (2.6) can be written as:

$$P(d)=Q\left\{\sum_{n=[\log_2 d]+1}^{d}\sum\nolimits_{(i_1,\ldots i_n)\in G(d,n)}\Lambda(i_{n-1},i_n)\cdots\Lambda(i_1,i_2)\right\}Q^{-1}P(1) \quad (2.11)$$

where: $\Lambda(i_{k-1},i_k)=\Xi(i_k)D(i_{k-1},i_k)$.

Denote by:

$$H(d)=\sum_{n=[\log_2 d]+1}^{d}\sum\nolimits_{(i_1,\ldots,i_n)\in G(d,n)}\Lambda(i_{n-1},i_n)\cdots\Lambda(i_1,i_2)$$

$$H(d)=\begin{bmatrix}H_1(d) & 0 & 0 & 0\\ 0 & H_2(d) & 0 & 0\\ 0 & 0 & H_2(d) & 0\\ 0 & 0 & 0 & H_3(d)\end{bmatrix}$$

where:

$$H_k(d)=\sum_{n=[\log_2 d]+1}^{d}\sum\nolimits_{(i_1,\ldots,i_n)\in G(d,n)}\delta_k(i_1,\ldots,i_n) \quad (2.12)$$

$$\delta_k(i_1,\ldots,i_n)=\frac{\pi_k(i_{n-1},i_n)\ldots\pi_k(i_1,i_2)}{(1-v(i_n))\ldots(1-v(i_2))} \quad (2.13)$$

**Proposition 2.11**: Let $\Theta = \cdots \alpha_{-1}\alpha_0\alpha_1\alpha_2 \cdots$ be an infinite string of binary symbols. Let us denote by $P_0(\Theta), P_1(\Theta)$ the densities of zeros and ones respectively in string $\Theta$. Then if $P_1(\Theta) \neq 1/2$ we have: $H_2(d) \equiv 1$.

**Proof**: We will use some properties of probabilities $P_{\alpha,\beta}(d)$ in strings of binary symbols (see for example [Li, 1990]): For every $d \geq 1$:

$$P_{0,1}(d) = P_{1,0}(d) \quad ; \quad P_{0,0}(d) = 1 - 2P_1(\Theta) + P_{1,1}(d) \quad ; \quad P_{1,0}(d) = P_1(\Theta) - P_{1,1}(d)$$

It is not difficult to see that:

$$\begin{bmatrix} P_{0,0}(d) \\ P_{0,1}(d) \\ P_{1,0}(d) \\ P_{1,1}(d) \end{bmatrix} = \frac{1}{4} M(P_1(\Theta), P_{1,1}(1)) \begin{bmatrix} H_1(d) \\ H_2(d) \\ H_3(d) \end{bmatrix} \qquad (2.14)$$

where:

$$M(P_1(\Theta), P_{1,1}(1)) = \begin{bmatrix} 1 & 2(1-2P_1(\Theta)) & 1-4P_1(\Theta)+4P_{1,1}(1) \\ 1 & 0 & -(1-4P_1(\Theta)+4P_{1,1}(1)) \\ 1 & 0 & -(1-4P_1(\Theta)+4P_{1,1}(1)) \\ 1 & -2(1-2P_1(\Theta)) & 1-4P_1(\Theta)+4P_{1,1}(1) \end{bmatrix}$$

Now because of $P_{00}(d) = 1 - 2P_1(\Theta) + P_{1,1}(d)$, from Eq. (2.14) we have:

$$(1-2P_1(\Theta))H_2(d) = 1 - 2P_1(\Theta)$$

This completes the proof.

From Eq. (2.1) of [Li, 1990] and Eq. (2.14) we have:

$$\Gamma(d) = H_1(d) + (1 - 4P_1(\Theta) + P_{1,1}(d))H_3(d) - (P_1^2(\Theta) - 4P_1(\Theta) + 2) \qquad (2.15)$$

where $\Gamma(d)$ is the correlation function. Hence we have obtained an expression for correlation function based on the eigenvalues of the matrix $M(i_{k-1}, i_k)$. Later we will obtain upper and lower bounds for $H_1(d)$ and $H_3(d)$.

## 3 Upper and Lower Bounds for Eigenvalues.

In this section we will obtain upper and lower bounds for $v_k(d)$ and $\pi_k(d^{'}, d)$ as functions of $d, d^{'}$ and $p$. First, we will give some definitions which can be seen in [Li, 1991a]. There, in Fig. 10 all the possible cases in which two binary symbols, previously separated at distance $d^{'}$ would be at distance $d$ in the next time step, are shown. These cases are labelled as $A_1$ , $A_2$ , $A_3$ , $B_1$ , $B_2$ and $C$. Their probabilities are (see Eq. B5 of Appendix B of [Li,1991a]):

$$P(A_1) = (1-p)^2 \binom{d^{'}-1}{2d^{'}-d+1} p^{2d^{'}-d+1}(1-p)^{d-d^{'}-2} \qquad (3.1a)$$

$$P(A_2) = (1-p)^2 \binom{d^{'}-1}{2d^{'}-d} p^{2d^{'}-d}(1-p)^{d-d^{'}-1} \qquad (3.1b)$$

$$P(A_3) = (1-p)^2 \binom{d^{'}-1}{2d^{'}-d-1} p^{2d^{'}-d-1}(1-p)^{d-d^{'}} \qquad (3.1c)$$

$$P(B_1) = p(1-p) \binom{d^{'}-1}{2d^{'}-d} p^{2d^{'}-d}(1-p)^{d-d^{'}-1} \qquad (3.1d)$$

$$P(B_2) = p(1-p) \binom{d^{'}-1}{2d^{'}-d-1} p^{2d^{'}-d-1}(1-p)^{d-d^{'}} \qquad (3.1e)$$

$$P(C) = p^2 \binom{d^{'}-1}{2d^{'}-d-1} p^{2d^{'}-d-1}(1-p)^{d-d^{'}} \qquad (3.1f)$$

The coefficients of matrix $T(d^{'},d,p)$ can be built in terms of them (see also Appendix B):

$$T_0(d^{'},d,p) = P(A_1) + 2P(A_2) + P(A_3) \qquad (3.2a)$$

$$T_1(d^{'},d,p) = P(B_1) + P(B_2) \qquad (3.2b)$$

$$T_2(d^{'},d,p) = P(C) \qquad (3.2c)$$

Now from the expressions (2.10), (3.2a), (3.2b), (3.2c) we have:

$$\pi_1(d^{'},d,p) = p^{2d^{'}-d-1}(1-p)^{d-d^{'}} \left\{ A(d^{'},d)p^2 + 2B(d^{'},d)p + C(d^{'},d) \right\} \qquad (3.3a)$$

$$\pi_2(d^{'},d,p) = p^{2d^{'}-d-1}(1-p)^{d-d^{'}} \left\{ A(d^{'},d)p^2 + 2B(d^{'},d)p(1-p) + C(d^{'},d)(1-2p) \right\}$$

(3.3b)

$$\pi_3(d^{'},d,p) = p^{2d^{'}-d-1}(1-p)^{d-d^{'}} \left\{ A(d^{'},d)p^2 + 2B(d^{'},d)p(1-2p) + C(d^{'},d)(1-2p)^2 \right\}$$

(3.3c)

where:

$$A(d^{'},d)=\binom{d^{'}-1}{2d^{'}-d+1}$$

$$B(d^{'},d)=\binom{d^{'}-1}{2d^{'}-d}$$

$$C(d^{'},d)=\binom{d^{'}-1}{2d^{'}-d-1}$$

It is not difficult to see that if $d^{'}=d$, then the only possible cases are $A_3, B_2$ y $C$. Therefore the eigenvalues of matrix the $T(d^{'},d^{'},p)$ are:

$$v_1(d^{'}) = p^{d^{'}-1} \tag{3.4a}$$

$$v_2(d^{'}) = p^{d^{'}-1}(1-2p) \tag{3.4b}$$

$$v_3(d^{'}) = p^{d^{'}-1}(1-2p)^2 \tag{3.4c}$$

**Theorem 3.1**: Let $\pi_1(d^{'},d,p), \pi_3(d^{'},d,p)$ be as defined by Eqs. (3.3a) y (3.3c). Then if $d^{'}$ is large enough and $0 \le p < 1/2$ we have:

$$\varphi_l^1(p)\frac{e^{-\frac{(d^{'}-1)(1-p)}{2p}}}{\sqrt{d^{'}-1}} \le \pi_1(d^{'},d,p) \le \varphi_u^1(p)\frac{e^{-\frac{(d^{'}-1)p}{2(1-p)}}}{\sqrt{d^{'}-1}} \tag{3.5a}$$

$$\varphi_l^3(p)\frac{e^{-\frac{(d^{'}-1)(1-p)}{2p}}}{\sqrt{d^{'}-1}} \le \pi_3(d^{'},d,p) \le \varphi_u^3(p)\frac{e^{-\frac{(d^{'}-1)p}{2(1-p)}}}{\sqrt{d^{'}-1}} \tag{3.5b}$$

where:

$$\varphi_l^1(p)=\frac{e^{\frac{1}{2(1-p)}}+(1-p)^2 e^{-\frac{3}{2p}}+2(1-p)}{\sqrt{2\pi p(1-p)}}$$

$$\varphi_u^1(p)=\frac{e^{\frac{1}{p}}+(1-p)^2 e^{\frac{3}{1-p}}+2(1-p)}{\sqrt{2\pi p(1-p)}}$$

$$\varphi_l^3(p) = \frac{e^{\frac{1}{2(1-p)}} + (1-2p)^2 e^{-\frac{3}{2p}} - 2(1-2p)}{\sqrt{2\pi p(1-p)}}$$

$$\varphi_u^3(p) = \frac{e^{\frac{1}{p}} + (1-2p)^2 e^{\frac{3}{1-p}} + 2(1-2p)}{\sqrt{2\pi p(1-p)}}$$

**Proof**:

Let $d_0 \in \mathrm{N}$ such that for every $d \geq d_0$, the approximation of Local Limit Theorem remain valid:

$$\binom{d}{r} p^r (1-p)^{d-r} \approx \frac{1}{\sqrt{2\pi p(1-p)d}} e^{-\frac{1}{2}\left(\frac{r-dp}{\sqrt{dp(1-p)}}\right)^2}$$

($d_0$ could be 25, see page 84 of [Gnedenko, 1980]). Then from (3.3a) , (3.3c) and under supposition that $d^{'} \geq d_0$, we have:

$$\pi_1 = \frac{\sqrt{w}}{\sqrt{2\pi p(1-p)}} e^{-\frac{u^2}{2wp(1-p)}} \left\{ e^{-\left(\frac{w+2u}{2(1-p)}\right)} + (1-p)^2 e^{-\left(\frac{w-2u}{2(1-p)}\right)} + 2(1-p) \right\} \qquad (3.7a)$$

$$\pi_3 = \frac{\sqrt{w}}{\sqrt{2\pi p(1-p)}} e^{-\frac{u^2}{2w(1-p)}} \left\{ e^{-\left(\frac{w-2u}{2(1-p)}\right)} + (1-2p)^2 e^{-\left(\frac{w+2u}{2(1-p)}\right)} - 2(1-2p) \right\} \qquad (3.7b)$$

where:

$$u = 1 - \frac{2d^{'} - d}{p(d^{'} - 1)} \qquad ; \qquad w = \frac{1}{p(d^{'} - 1)}$$

From condition c) of Definition 2.1 we have $d^{'} + 1 \leq d \leq 2d^{'} + 1$, hence:

$$-\frac{1}{p(d^{'} - 1)} \leq \frac{2d^{'} - d}{p(d^{'} - 1)} \leq \frac{1}{p}$$

and then:

$$e^{-\frac{3}{2p}} \leq e^{-\left(\frac{w-2u}{2(1-p)}\right)} \leq e^{\frac{3}{1-p}}$$

$$e^{\frac{1}{2(1-p)}} \le e^{-\left(\frac{w+2u}{2(1-p)}\right)} \le e^{\frac{1}{p}}$$

From the above inequalities and (3.7a) and (3.7b) we could obtain:

$$\varphi_l^1(p)\frac{e^{-\frac{u^2}{2w(1-p)}}}{\sqrt{d'-1}} \le \pi_1(d',d,p) \le \varphi_u^1(p)\frac{e^{-\frac{u^2}{2w(1-p)}}}{\sqrt{d'-1}} \qquad (3.8a)$$

$$\varphi_l^3(p)\frac{e^{-\frac{u^2}{2w(1-p)}}}{\sqrt{d'-1}} \le \pi_3(d',d,p) \le \varphi_u^3(p)\frac{e^{-\frac{u^2}{2w(1-p)}}}{\sqrt{d'-1}} \qquad (3.8b)$$

Let us note that:

$$\frac{u^2}{2w(1-p)} = \frac{p(d'-1)}{2(1-p)}\left(1-\frac{2d'-d}{p(d'-1)}\right)^2$$

It is not difficult to prove that if $d'+1 \le d \le 2d'+1$, then:

$$1 \le \left|1-\frac{2d'-d}{p(d'-1)}\right| \le \frac{1-p}{p}$$

The above condition , (3.8a) and (3.8b) complete the proof.

**Remark**: It is easy to see that $\varphi_l^3(p) \ge 0$ if $0.0716 \le p$. In all that follows we suppose that $p$ satisfies the above mentioned condition.

**Definition 3.2**: Let $(i_1,\ldots,i_n) \in G(d,n)$ and $d_0 \in \mathrm{N}$. Let us denote by $l(i_1,\ldots,i_n)$ the set of indexes which are smaller than $d_0$ and by $u(i_1,\ldots,i_n)$ those which are bigger:

$$l(i_1,\ldots,i_n) = \{i_r \in (i_1,\ldots,i_n): i_r < d_0\}$$

$$u(i_1,\ldots,i_n) = \{i_r \in (i_1,\ldots,i_n): i_r \ge d_0\}$$

**Definition 3.3**: Let $(i_1,\ldots,i_n) \in G(d,n)$ . Denote by:

$$c_j(i_1,\ldots,i_n) = \prod_{l(i_1,\ldots,i_n)} \frac{\pi_j(i_k,i_{k+1},p)}{1-v(i_{k+1})} \qquad j=1,3$$

**Corollary 3.4**: Under conditions of Theorem 3.1 we have:

$$L_j(p,d_0,d,n) \le \delta_j(i_1,\ldots,i_n) \le U_j(p,d_0,d,n) \qquad (3.9)$$

where:

$$L_j(p,d_0,d,n) = c_j(i_1,\ldots,i_n)\Phi_l^j(p,d_0,d,n)\frac{e^{-\frac{1-p}{2p}S(d,n)}}{M(d,n)} \tag{3.10a}$$

$$U_j(p,d_0,d,n) = c_j(i_1,\ldots,i_n)\Phi_u^j(p,d_0,d,n)\frac{e^{-\frac{p}{2(1-p)}S(d,n)}}{M(d,n)} \tag{3.10b}$$

$$\Phi_l^j(p,d_0,d,n) = \prod_{l(i_1,\ldots,i_n)} \frac{\varphi_l^j(p)}{1-v_j(i_k)}$$

$$\Phi_u^j(p,d_0,d,n) = \prod_{l(i_1,\ldots,i_n)} \frac{\varphi_u^j(p)}{1-v_j(i_k)}$$

$$S(d,n) = \sum_{u(i_1,\ldots,i_n)} (i_k - 1) \quad ; \quad M(d,n) = \prod_{u(i_1,\ldots,i_n)} (i_k - 1)$$

**Proof**: The expression $\delta_j(i_1,\ldots,i_n)$ can be written:

$$\delta_j(i_1,\ldots,i_n) = \prod_{l(i_1,\ldots,i_n)} \frac{\pi_j(i_k,i_{k+1})}{1-v_j(i_{k+1})} . \prod_{u(i_1,\ldots,i_n)} \frac{\pi_j(i_k,i_{k+1})}{1-v_j(i_{k+1})}$$

therefore:

$$\delta_j(i_1,\ldots,i_n) = c_j(i_1,\ldots,i_n) \prod_{u(i_1,\ldots,i_n)} \frac{\pi_j(i_k,i_{k+1})}{1-v_j(i_{k+1})}$$

Now from (3.5a) and (3.5b) we have:

$$\Phi_l^j(p,d_0,d,n)\frac{e^{-\frac{1-p}{2p}S(d,n)}}{M(d,n)} \le \prod_{u(i_1,\ldots,i_n)} \frac{\pi_j(i_k,i_{k+1})}{1-v_j(i_{k+1})}$$

$$\Phi_u^j(p,d_0,d,n)\frac{e^{-\frac{p}{2(1-p)}S(d,n)}}{M(d.n)} \ge \prod_{u(i_1,\ldots,i_n)} \frac{\pi_j(i_k,i_{k+1})}{1-v_j(i_{k+1})}$$

The above inequalities complete the proof.

## 4 Upper and Lower Bounds for Sum and Product of Indexes.

In Corollary 3.4 we obtain upper and lower bounds for $\delta_j(i_1,\dots,i_n)$. Those expressions depend on $S(d,n)$ and $M(d,n)$. In this section we will obtain upper and lower bounds for $S^*(d,n)=\sum_{(i_1,\dots,i_n)\in G(d,n)}(i_k-1)$ and $M^*(d,n)=\prod_{(i_1,\dots,i_n)\in G(d,n)}(i_k-1)$, which together with (3.10a) y (3.10b) will be used to obtain uniform bounds for $\delta_j(i_1,\dots,i_n)$ on $G(d,n)$.

The expression $S^*(d,n)$ as well as $M^*(d,n)$ reach their maximum values in those members of the set $G(d,n)$ in which the last elements are consecutive, e.g., $i_n=d$, $i_{n-1}=d-1$, $i_{n-2}=d-2$, etc. This condition constrains to the first ones to be as sparse as possible, but fulfilling the condition: $[i_{k+1}/2]=i_k$, e.g.:

$$(i_1,i_2,\dots,i_{n-r-1},i_{n-r},\dots,i_n)=(1,3,\dots,2^{n-r-1}-1,d-r,\dots,d)$$

Therefore we should find an index $r$ such that:

$$\begin{cases} \left[\dfrac{d-r}{2}\right]\le 2^{n-r-1}-1 \\ 2^{n-r-2}-1<\left[\dfrac{d-r-1}{2}\right] \end{cases} \tag{4.1}$$

If $d-r$ is even, the above conditions are equivalent to:

$$\begin{cases} d+2\le 2^{n-r}+r \\ 2^{n-(r+1)}+r+1\le d \end{cases} \tag{4.2}$$

If $d-r$ is odd, conditions (3.11) are equivalent to:

$$\begin{cases} d+1\le 2^{n-r}+r \\ 2^{n-(r+1)}+r+1\le d+1 \end{cases} \tag{4.3}$$

In order to obtain such an index $r$ we will solve the equation:

$$2^{n-x}+x=d+1 \tag{4.4}$$

In Fig. 4.1 it is shown the graph of difference between solutions of equations:

$$2^{n-x}+x=d \qquad ; \qquad 2^{n-x}+x=d+2$$

as function of $d$ . This justifies the search of index $r$ using (4.4).

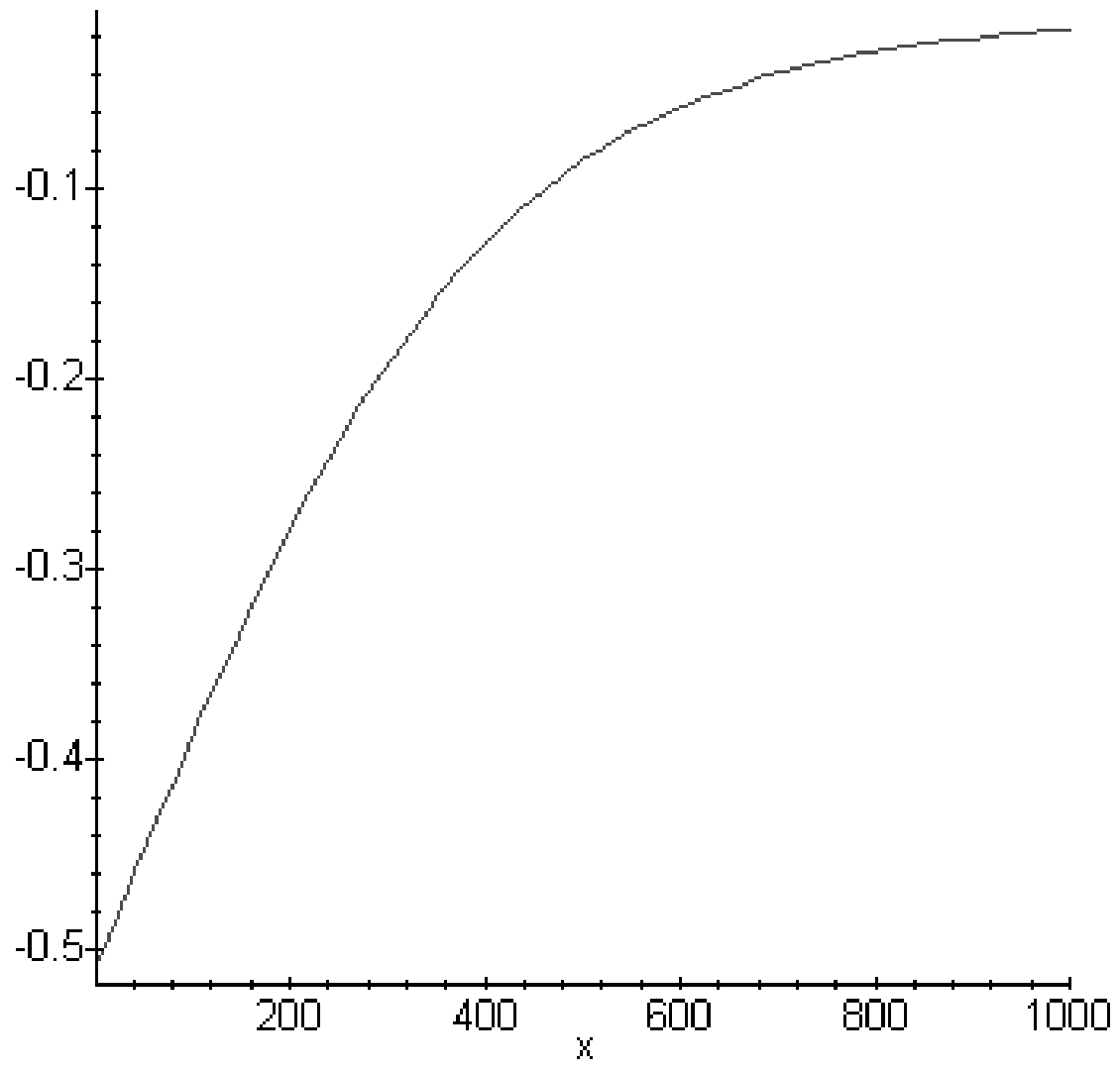

Fig. 4.1: The difference between the solutions of $2^{n-x}+x=d$ and $2^{n-x}+x=d+2$. In the x axis is shown the distance $d$ .

It can be shown that the only feasible solution of (4.4) can be expressed by means of the following series expansion:

$$x=\sum_{k=1}^{\infty}\frac{1}{k!}\frac{(\ln 2)^k P_k(d)}{(d\ln 2-1)^{2k-1}}\left(n-\frac{\ln d}{\ln 2}\right)^k \tag{4.5}$$

where $P_k(d)$ is a polynomial in $d$ of degree $k-1$. The expression $n-\frac{\ln d}{\ln 2}$ is always positive because $n \geq [\log_2 d]+1$.

The above series could be written as:

$$x=\sum_{k=1}^{\infty}\frac{1}{k!}\frac{P_k(d)}{(d\ln 2-1)^{k-1}}\left(\frac{n\ln 2-\ln d}{d\ln 2-1}\right)^k$$

Let us note that:

$$0 \leq \frac{n\ln 2-\ln d}{d\ln 2-1} \leq \frac{d\ln 2-\ln d}{d\ln 2-1}$$

and it is easy to see that:

$$\lim_{d \to \infty} \frac{P_k(d)}{(d \ln 2 - 1)^k} = \frac{1}{\ln 2}$$

Therefore, as solution of (4.4) we will use the approximation done by the first term of series (4.5):

$$r_u = \frac{d \ln 2}{d \ln 2 - 1}\left(n - \frac{\ln d}{\ln 2}\right) \qquad (4.6)$$

We want to remark the accuracy of this approximation. In Table II are shown some couples of $(d,n)$, the exact values of $n-r$ and the approximation obtained from (4.6) (labelled $n-r_u$).

From the result obtained above we could have upper bound for $S^*(d,n)$ *y* $M^*(d,n)$:

$$S^*(d,n) \leq S_u(d,n) \qquad \text{y} \qquad M^*(d,n) \leq M_u(d,n) \qquad (4.7)$$

where:

$$S_u(d,n) = d^{1-\frac{n}{d \ln d}} + 2d(n + 2 - \frac{\ln d}{\ln 2}) + \frac{\ln d}{\ln 2}(2n - 1 + \frac{\ln d}{\ln 2}) \qquad (4.8)$$

$$M_u(d,n) = 2^{\frac{(n-r_u)(n-r_u-1)}{2}} \frac{d!}{(d - r_u - 1)!} \qquad (4.9)$$

## TABLE II
## Exact values of $n-r$ and their estimates $n-r_u$

| $d$ | $n$ | $n-r$ | $n-r_u$ |
|---|---|---|---|
| **30** | **10** | **5** | **4.64** |
| **30** | **20** | **4** | **4.14** |
| **33** | **11** | **5** | **4.77** |
| **60** | **20** | **6** | **5.55** |
| **60** | **40** | **5** | **5.06** |
| **62** | **7** | **6** | **5.92** |
| **80** | **13** | **7** | **6.19** |
| **81** | **18** | **7** | **6.12** |
| **120** | **40** | **7** | **6.50** |
| **120** | **80** | **6** | **6.01** |
| **240** | **80** | **8** | **7.47** |
| **240** | **160** | **7** | **6.98** |

We proceed in the same way for lower bound. The expression $S^*(d,n)$ as well as $M^*(d,n)$ reach their minimum values in those members of the set $G(d,n)$ in which the last elements are as sparse as possible. This condition constrains the first ones to be consecutive, because the condition $i_k < i_{k+1}, k = 1,\dots,n-1$ must be fulfilled.

$$(i_1,\dots,i_r,i_{r+1},\dots,i_n) = (1,\dots,r,i_{r+1},\dots,i_n)$$

where: $\left[\dfrac{i_{l+1}}{2}\right] \le i_l$ ; $l = r+1,\dots,n-1$.

Therefore, we should find an index $r$ such that:

$$\begin{cases} 2^{n-(r+1)}(r+1) < d+1 \\ d+1 < 2^{n-r}r \end{cases} \qquad (4.10)$$

To obtain such an index, we will solve the equation:

$$2^x (n - x) = d + 1 \qquad (4.11)$$

It can be proved that the only feasible solution of (4.11) is expressed by means of the following series expansion:

$$x = \sum_{k=1}^{\infty} \frac{1}{k!} \frac{(\ln 2)^k P_k(d)}{(d \ln 2 - 1)^{2k-1}} (d - n)^k \qquad (4.12)$$

where $P_k(d)$ is a polynomial of degree $k - 1$. More precisely:

$$P_k(d) = (k-1)!(\ln 2)^{k-2} d^{k-1} + \ldots + d$$

Besides, it is not difficult to see that:

$$0 \le \frac{d \ln 2 - n \ln 2}{d \ln 2 - 1} \le \frac{d \ln 2 - \ln d}{d \ln 2 - 1} \le 1$$

and:

$$\lim_{d \to \infty} \frac{P_k(d)}{(d \ln 2 - 1)^k} = \frac{1}{\ln 2}$$

Hence, for $d$ large enough:

$$x \approx \frac{1}{\ln 2} \sum_{k=1}^{\infty} \frac{1}{k} \left( \frac{d \ln 2 - n \ln 2}{d \ln 2 - 1} \right)^k = \frac{1}{\ln 2} \ln \left[ \frac{d \ln 2 - 1}{n \ln 2 - 1} \right] \qquad (4.13)$$

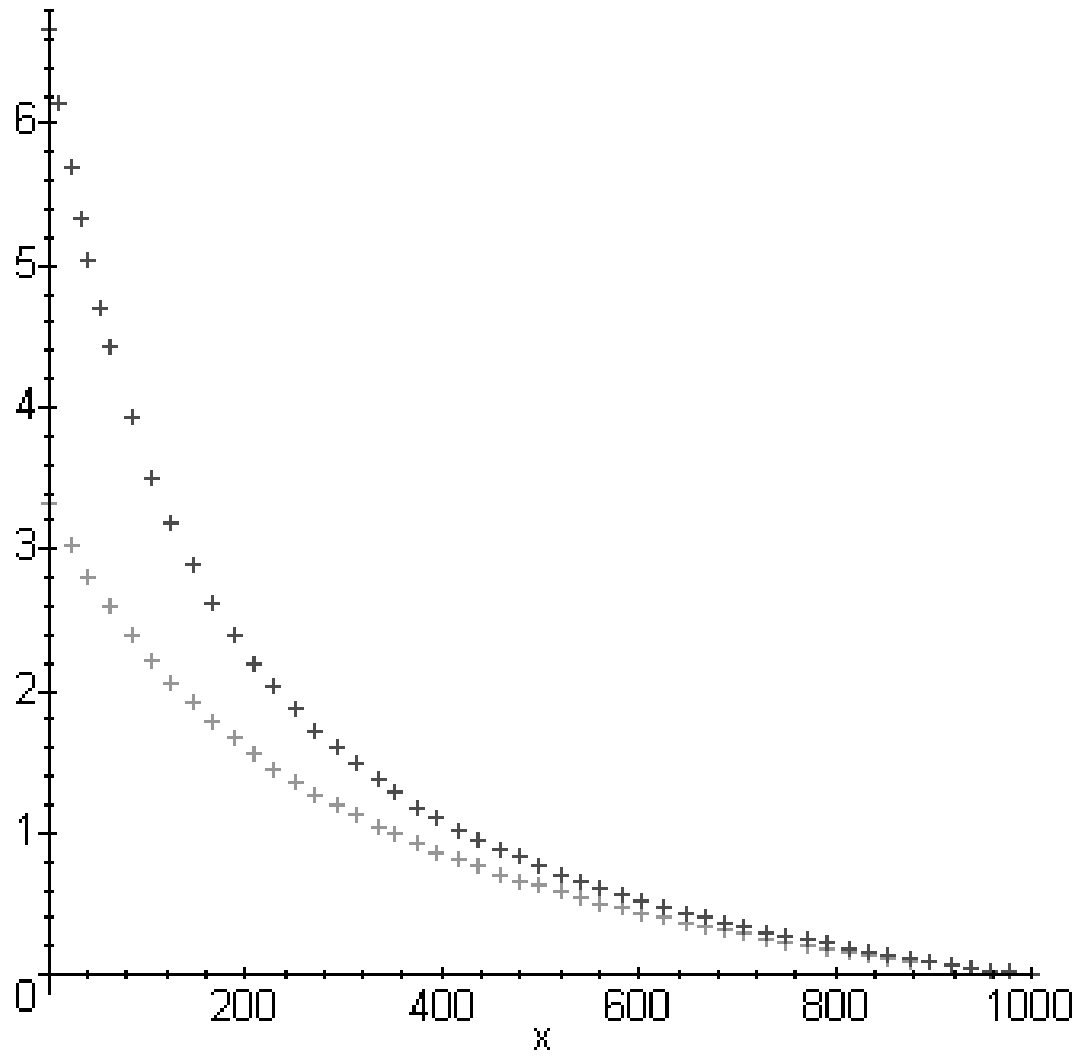

Fig 4.2: The graphs of the whole series (4.12) and the approximation (4.13) for n=20. In the x axis is shown the distance $d$ .

We will take as approximation to solution of (4.11) the expression (4.13). In Fig. 4.2 we present the graphs of the whole series (upper plot) and the approximation (4.13) (lower plot) for $d = 1024$ and $11 \le n \le 1000$. In Table III we present some couples of $(d, r)$, the corresponding values of $r$ and the approximation obtained from (4.13), ( labelled $r_l$ ).

## TABLE III
## Exact values of de $n - r$ and their estimates $n - r_l$.

| d | n | $n - r$ | $n - r_l$ |
|---|---|---|---|
| **30** | **10** | **9** | **8.26** |
| **30** | **20** | **20** | **19.37** |
| **33** | **11** | **11** | **9.27** |
| **60** | **20** | **19** | **18.34** |
| **60** | **40** | **40** | **39.39** |
| **62** | **7** | **4** | **3.55** |
| **80** | **13** | **11** | **10.23** |
| **81** | **18** | **16** | **15.73** |
| **120** | **40** | **39** | **38.77** |
| **120** | **80** | **80** | **79.40** |
| **240** | **80** | **79** | **78.34** |
| **240** | **160** | **160** | **159.41** |

From all above:

$$S^*(d,n) \geq S_l(d,n) \quad ; \quad M^*(d,n) \geq M_l(d,n) \qquad (4.14)$$

where:

$$S_l(d,n) = \frac{(n-r_l)(n-r_l-1)-2n}{2} + (d+1)\left\{\frac{(d-n)\ln 2 - 1}{2d\ln 2 - 1}\right\} - \frac{1}{\ln 2}\ln\left[\frac{d\ln 2 - 1}{n\ln 2 - 1}\right] + 1 \quad (4.15)$$

$$M_l(d,n) = d(n-r_l-1)!(d+1)^{r_l}\, 2^{-\frac{r_l(r_l+1)}{2}}\, e^{\frac{2}{d+1}\left(2^{r_l}-1\right)} \qquad (4.16)$$

## 5 Uniform Bounds on Sets G(d,n).

In this section, starting from (4.15) and (4.16), we will obtain upper and lower bounds for $\delta_j(i_1,\ldots,i_n)$ as functions of $d$ and $n$ on every set $G(d,n)$.

**Definition 5.1**: Let us denote for:

$$c_j^u(d_0,n) = \max_{(i_1,\ldots,i_n)\in G(d,n)} c_j(i_1,\ldots,i_n)$$

$$c_j^l(d_0,n) = \min_{(i_1,\ldots,i_n)\in G(d,n)} c_j(i_1,\ldots,i_n)$$

**Proposition 5.2**:

The following inequalities hold:

$$L_j(p,d_0,d,n) \geq c_j^l(d_0,n)\Phi_l^j(p,d_0,d,n)Q^l(d_0,n)\frac{e^{-\frac{1-p}{2p}S_u(d,n)}}{M_u(d,n)} \qquad (5.1)$$

$$U_j(p,d_0,d,n) \leq c_j^u(d_0,n)\Phi_u^j(p,d_0,d,n)Q^u(d_0,n)\frac{e^{-\frac{p}{2(1-p)}S_l(d,n)}}{M_l(d,n)} \qquad (5.2)$$

where:

$$Q^l(d_0,n) = \min_{(i_1,\ldots,i_n)\in G(d,n)} W^l(i_1,\ldots,i_n)$$

$$Q^u(d_0,n) = \max_{(i_1,\ldots,i_n)\in G(d,n)} W^u(i_1,\ldots,i_n)$$

$$W^l(i_1,\ldots,i_n) = \prod_{l(i_1,\ldots,i_n),\,k\neq 1} \sqrt{i_k - 1}\; e^{\frac{1-p}{2p}\sum_{l(i_1,\ldots,i_n)}(i_k - 1)}$$

$$W^u(i_1,\ldots,i_n) = \prod_{l(i_1,\ldots,i_n),\,k\neq 1} \sqrt{i_k - 1}\; e^{\frac{p}{2(1-p)}\sum_{l(i_1,\ldots,i_n)}(i_k - 1)}$$

**Proof**: From (3.10a) and (3.10b) we have:

$$L_j(p,d_0,d,n) \geq c_j^l(d_0,n)\Phi_l^j(p,d_0,d,n)W^l(i_1,\ldots,i_n)\frac{e^{-\frac{1-p}{2p}S^*(d,n)}}{M^*(d,n)}$$

$$U_j(p,d_0,d,n) \leq c_j^u(d_0,n)\Phi_u^j(p,d_0,d,n)W^u(i_1,\ldots,i_n)\frac{e^{-\frac{p}{2(1-p)}S^*(d,n)}}{M^*(d,n)}$$

From the above inequalities and expressions (4.7) and (4.14) we have (5.1) and (5.2).

**Proposition 5.3**: If $0.08 \leq p \leq 0.25$, then for $d$ large enough and $j = 1,3$:

$$\Phi_u^j(p,d_0,d,n) \leq e^{n\left(2\ln 2 + \frac{1}{p} + \frac{p^{d_0}}{1-p}\right)}(2\pi(1-p)p)^{-\frac{n}{2}} = B_u(p,d_0,n) \quad (5.3)$$

$$\Phi_l^j(p,d_0,d,n) \geq (2\pi(1-p)p)^{-\frac{1}{2}}\, e^{\frac{1}{2(1-p)} - \ln 5} = B_l(p,d_0,n) \quad (5.4)$$

**Proof**: If $p < 0.25$, we could prove that.

$$\varphi_u^j(p) \leq 4e^{\frac{1}{4}}(2\pi(1-p)p)^{-\frac{1}{2}}$$

Besides, from:

$$1 - p^k = e^{\ln(1-p^k)} = e^{-p^k + o(p^k)}$$

we have, for $d$ large enough:

$$\prod_{u(i_1,\ldots,i_n)}(1 - p^{i_k}) \geq e^{-(p^{d_0} + \ldots + p^d) + o(p^{d_0})} = e^{-\frac{p^{d_0}}{1-p} + o(p^{d_0})}$$

Therefore:

$$\Phi_u^j(p,d_0,d,n) \le e^{n\left(2\ln 2+\frac{1}{p}+\frac{p^{d_0}}{1-p}\right)}(2\pi(1-p)p)^{-\frac{n}{2}}$$

If $0.08 < p$, then:

$$\frac{1}{5}(2\pi(1-p)p)^{\frac{1}{2}} e^{\frac{1}{2(1-p)}} \le \varphi_l^j(p)$$

and from the above:

$$(2\pi(1-p)p)^{-\frac{1}{2}} e^{\frac{1}{2(1-p)}-\ln 5} \le \Phi_l^j(p,d_0,d,n)$$

This completes the proof.

**Proposition 5.4**: Under the same conditions of Proposition 5.3 we have:

$$\sum_{(i_1,\ldots,i_n)\in G(d,n)} \delta_j(i_1,\ldots,i_n) \le K^u(p,d_0,n)\frac{e^{-\frac{p}{2(1-p)}S_l(d,n)}}{M_l(d,n)}\frac{(2d-n-3)}{n-3}\binom{d-4}{n-4} \quad (5.5)$$

$$\sum_{(i_1,\ldots,i_n)\in G(d,n)} \delta_j(i_1,\ldots,i_n) \ge K^l(p,d_0,n)\frac{e^{-\frac{1-p}{2p}S_u(d,n)}}{M_u(d,n)} \quad (5.6)$$

where:

$$K^u(p,d_0,n) = c_j^u(d_0,n)B_u(p,d_0,n)Q^u(d_0,n)$$

$$K^l(p,d_0,n) = c_j^l(d_0,n)B_l(p,d_0,n)Q^l(d_0,n)$$

**Proof**:

Let $(i_1,\ldots,i_n) \in G(d,n)$. Then from Corollary 3.4 and Propositions 5.2 , 5.3 we have:

$$\delta_j(i_1,\ldots,i_n) \le K^u(p,d_0,n)\frac{e^{-\frac{p}{2(1-p)}S_l(d,n)}}{M_l(d,n)} \quad (5.7)$$

$$\delta_j(i_1,\ldots,i_n) \geq K^l(p,d_0,n)\frac{e^{-\frac{1-p}{2p}S_u(d,n)}}{M_u(d,n)} \tag{5.8}$$

From (5.8) we obtain (5.6) straightforward. From (5.7) and expression (A.9) of Appendix A we obtain (5.5). This completes the proof:

The expressions (5.5) and (5.6) can be written as:

$$\sum_{(i_1,\ldots,i_n)\in G(d,n)} \delta_j(i_1,\ldots,i_n) \geq c_j^l(d_0,n)Q^l(d_0,n)\, e^{-\varepsilon_l(d,n,p)} \tag{5.9}$$

$$\sum_{(i_1,\ldots,i_n)\in G(d,n)} \delta_j(i_1,\ldots,i_n) \leq c_j^u(d_0,n)Q^u(d_0,n)\, e^{-\varepsilon_u(d,n,p)} \tag{5.10}$$

where:

$$\varepsilon_u(d,n,p) = \varepsilon_u^1(d,n,p) + \varepsilon_u^2(d,n,p) + \varepsilon_u^3(d,n,p) + \varepsilon_u^4(d,n,p) \tag{5.11}$$

$$\varepsilon_u^1(d,n,p) = \frac{p}{2(1-p)}(d+1)\left[\frac{(d-n)\ln 2 - 1}{d\ln 2 - 1}\right] +$$

$$+\frac{1}{\ln 2}\ln\left[\frac{d\ln 2 - 1}{n\ln 2 - 1}\right]\left\{\ln(d+1) - \frac{1}{2}\left[\ln\left(\frac{d\ln 2 - 1}{n\ln 2 - 1}\right) + \ln 2\right] - \frac{p}{1-p}\right\}$$

$$\varepsilon_u^2(d,n,p) = (n - r_l - 1)\left\{\frac{n - r_l}{2} + \ln(n - r_l - 1) - 1\right\} + \frac{p}{1-p}(1-n) +$$

$$+(d-n)\left\{\ln(d-n) + \frac{2}{d+1}\left(\frac{\ln 2}{n\ln 2 - 1}\right)\right\}$$

$$\varepsilon_u^3(d,n,p) = \ln d + \frac{1}{2}\ln(n - r_l - 1) + \frac{\ln(d-n)}{2} + \ln(n-4)(n - 7/2) + \ln(n-3)$$

$$\varepsilon_u^4(d,n,p) = -\left\{\frac{\ln(d-4)}{2} + (d-4)\ln(d-4) + \ln(2d - n - 3) + n\left(2\ln 2 + \frac{1}{p} + \frac{p^{d_0}}{1-p}\right)\right\}$$

$$\varepsilon_l(d,n,p) = \varepsilon_l^1(d,n,p) + \varepsilon_l^2(d,n,p) + \varepsilon_l^3(d,n,p) \tag{5.12}$$

$$\varepsilon_l^1(d,n,p) = \frac{1-p}{p}\left\{d^{1-\frac{n}{d\ln d}} + 2d\left(n + 2 - \frac{\ln d}{\ln 2}\right) + \frac{\ln d}{\ln 2}\left(2n - 1 + \frac{\ln d}{\ln 2}\right)\right\}$$

$$\varepsilon_l^2(d,n,p) = \frac{\ln 2}{4}(n - r_u)(n - r_u - 1) + \frac{\ln d}{2} + \frac{d\ln d}{2} + (d - r_u - 1)$$

$$\varepsilon_l^3(d,n,p) = -\frac{1}{2}\left\{d + \ln(d - r_u - 1) + (d - r_u - 1)\ln(d - r_u - 1)\right\}$$

It can be proved that when $d \to \infty$ the following asymptotic expansions hold:

$$\varepsilon_u(d,n,p) = \varepsilon_u^a(d,n,p) + o\left(\frac{1}{d}, \frac{1}{n}\right) \tag{5.13}$$

$$\varepsilon_l(d,n,p) = \varepsilon_l^a(d,n,p) + o\left(\frac{1}{d}, \frac{1}{n}\right) \tag{5.14}$$

where:

$$\varepsilon_u^a(d,n,p) = \frac{pd}{4(1-p)} + \frac{n^2}{2} + n\left[2\ln n + \left(1 + \frac{1}{\ln 2}\right)(\ln n - \ln d) - \frac{5-2p}{2(1-p)}\right] +$$
$$+\frac{1}{2}\left(1 + \frac{1}{\ln 2}\right)\left(\frac{\ln d}{\ln 2}\right)^2$$

$$\varepsilon_l^a(d,n,p) = \frac{d}{2}\left[1 + (2n+5)\left(\frac{1-p}{p}\right)\right] + \frac{1}{2}\frac{\ln d}{\ln 2}\left[n + \frac{1-p}{p}\left(1 - \frac{d}{2}\right) - \frac{\ln 2}{2}\right] + \frac{1}{2}(2n-1)\frac{1-p}{p}$$

Now from (2.11) and expressions (5.9) and (5.10) we have:

$$H_j(d) \le \sum_{n=[\log_2 d]+1}^{d} c_j^u(d_0,n) Q^u(d_0,n)\, e^{-\varepsilon_u(d,n,p)} \tag{5.15}$$

$$H_j(d) \ge \sum_{n=[\log_2 d]+1}^{d} c_j^l(d_0,n) Q^l(d_0,n)\, e^{-\varepsilon_l(d,n,p)} \tag{5.16}$$

From the above inequalities and Eq. (2.15) upper and lower bounds for correlation function can be obtained.

## 6 The Exponent of the Major Contributing Term.

In this section we prove that certain set $G(d,n)$ makes a major contribution to the function $H_j(d)$. We also prove the uniqueness of such a set $G(d,n)$.

If the symbols $\alpha$, $\beta$ are at distance $d'$, then between them there are $d' - 1$ other symbols. Hence, in the next time step we will have $p(d'-1) + (1-p)(d'-1) = (d'-1)(2-p)$ symbols in average between the offspring of $\alpha$ and $\beta$. Taking into account the six possible cases mentioned in Sec. 3, it is easy to prove that the expected values for the distance is $\bar{d} = (d' - 1)(2 - p) + 2$. Therefore, for a certain value of $n$ there exists an element $(i_1, \ldots, i_n) \in G(d, n)$, such that, for every $k = 1, \ldots, n-1$: $i_{k+1} = \left[(2-p)(i_k - 1) + 2\right]$.

**Definition 6.1**: Let $d \in \mathrm{N}$ and $0 < p < 1$. Let $n(d,p)$ be the positive integer such that the set $G(d, n(d,p))$ contains the element $(i_1, \ldots, i_{n(d,p)})$ which holds the following condition: for every $k = 1, \ldots, n(d,p) - 1$: $i_{k+1} = \left[(2-p)(i_k - 1) + 2\right]$.

**Proposition 6.2**: For $p \to 0$ and $d \to \infty$ we have:

$$n(d,p) = 1 + \left[\frac{\ln(d(1-p)+p)}{\ln(2-p)}\right] \qquad (6.1)$$

**Proof**: Although $i_{k+1} = \left[(i_k - 1)(2-p) + 2\right]$, we have:

$i_{k+1} = (2-p)i_k + p + \varepsilon_k$ , where: $0 \le \varepsilon_k < 1$. It can be proved by induction that:

$$i_k = (2-p)^{n-1} + p\left\{\frac{(2-p)^{n-1} - 1}{1-p}\right\} + \varepsilon_1 (2-p)^{n-2} + \cdots + \varepsilon_{k-1}$$

in particular for $k = n(d,p)$:

$$d = (2-p)^{n(d,p)-1} + p\left\{\frac{(2-p)^{n(d,p)-1} - 1}{1-p}\right\} + \varepsilon_1 (2-p)^{n(d,p)-2} + \cdots + \varepsilon_{n(d,p)-1}$$

From the above equation and the condition imposed to $\varepsilon_i$ we have:

$$n(d,p) \le 1 + \frac{\ln(d(1-p)+p)}{\ln(2-p)} < n(d,p) + \frac{1}{\ln(2-p)}\left\{\ln 2 + \ln\left(1 - \frac{1}{2(2-p)^{n(d,p)-1}}\right)\right\}$$

From the last expression and for $d \to \infty$ , $p \to 0$ we obtain (6.1).

**Remark**: Let us note that:

$$\lim_{p \to 0} n(d,p) = 1 + \left[\frac{\ln d}{\ln 2}\right] = 1 + \left[\log_2 d\right]$$

$$\lim_{p \to 1} n(d,p) = d$$

in agreement with the fact: $1 + \left[\log_2 d\right] \le n \le d$ .

# 7 Discussion.

In Fig. 7.1 we show the graph of $\varepsilon_u^a(d, n(d, p), p)$ for $p = 0.1$ (lower plot) and that obtained from averaging 10 simulations with the same values of $p$ (upper plot). We want to remark the coincidence in shape of both plots. In [Chatzidimitriou-Dreismann & Larhmar, 1993] a magnitude inversely proportional to correlation function is studied. For that magnitude a fitness of the form $F(d) = d^{\varphi(d)}$ is obtained. In Fig. 1 of that paper a plot of the exponent $\varphi(d)$ is shown. We also want to emphasise the coincidence in shape of that exponent and the lower plot in Fig. 7.1 of our present work.

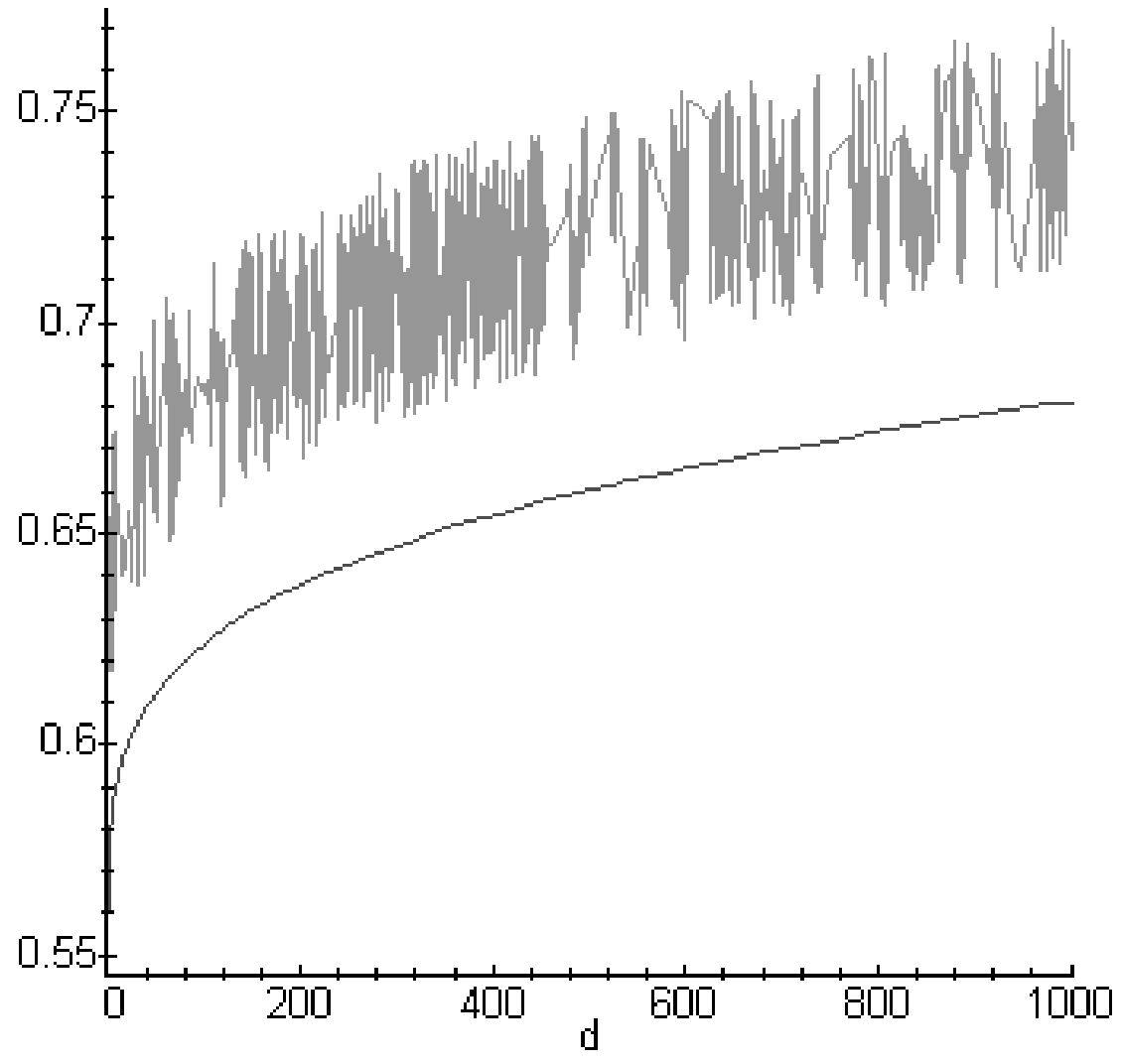

Fig. 7.1: The graph of $\varepsilon_u^a(d, n(d, p), p)$ for $p = 0.1$ (lower plot) and experimental exponent obtained from averaging 10 simulations with the same value of p. Compare with Fig. 1 of [Chatzidimitriou-Dreismann & Larhmar,1993].

In Fig. 1 of [Buldyrev et al., 1995] the graph of the averaged power spectra for all 33301 coding and for all 29453 noncoding sequences of the GENBANK larger than 512 bp is shown. As the authors remark, there are three spectral regimes. In our opinion, the existence of several exponents is the best explanation for that behaviour.

## 8 Conclusion.

We have studied the correlation function of an expansion-modification system which grasp the main features of the mutational process occurring in the evolution of DNA molecule. We obtain bounds for the exponent in the correlation function and show the resembling between the theoretical exponent and those obtained from simulation. We also give an explanation for the existence of several region in power spectra of real sequences.

The authors are thankful to G. Martinez-Meckler, R. Bujalich, P. Miramontes for their helpful comments and Carlos Gonzalez chairman of Digintec Corp for his valuable sofware support. The authors also thank to an anonymuos referee for his useful comments. This work was partially (R. Mansilla) supported by DGAPA-UNAM, Mexico.

**Appendix A: The Proof of Inequality (2.5).**

**Lemma A.1**: Under the same conditions of Lemma 2.3, let S be the set:

$$S = \{(n,d) \in \mathrm{N}^2 : d > n > 1, [d/2] \ge n,\ d\ even\}$$

then:

$$\theta(d,n+1) = \begin{cases} \theta(d-1,n) + \theta(d-1,n+1) - \theta(d/2-1,n) & (n,d) \in S \\ \theta(d-1,n) + \theta(d-1,n+1) & (n,d) \notin S \end{cases}$$

**Proof**:

We analyse several cases:

I- If $u(d,n) = d-1$, then:

$$\theta(d,n+1) = \sum_{k=l(d,n)}^{d-2} \theta(k,n) + \theta(d-1,n) \qquad \text{(A.1)}$$

I.1- If $l(d,n) = n$; then from $u(d,n) = d - 1$, it follow straightforward: $u(d-1,n) = d-2$ and because: $l(d-1,n) = l(d,n)$ we have:

$$\theta(d-1,n+1) = \sum_{k=l(d,n)}^{d-2} \theta(k,n) \qquad \text{(A.2)}$$

Now from (A.1) we have:

$$\theta(d,n+1) = \theta(d-1,n+1) + \theta(d-1,n) \qquad \text{(A.3)}$$

I.2- If $l(d,n) = [d/2]$; then it is not difficult to see that:

$$[(d-1)/2] = \begin{cases} d/2-1 & d \quad is \quad even \\ [d/2] & d \quad is \quad odd \end{cases} \qquad \text{(A.4)}$$

From the above it follow that if $d$ is odd: $l(d,n) = l(d-1,n)$ and therefore we obtain (A.2) and (A.3).

If $d$ is even , we analyse two cases:

I.2.1- If $d/2 = n$; then $d/2 - 1 < n$ and:

$$l(d-1,n) = n = d/2 = l(d,n)$$

from the above equation we obtain (A.2) and (A.3).

I.2.2- If $d/2 - 1 \geq n$; then $l(d-1,n) = d/2 - 1$ therefore: $l(d-1,n) = l(d,n) - 1$; from the last equation we have:

$$\theta(d-1,n+1) - \theta(d/2-1,n) = \sum_{k=l(d,n)}^{d-2} \theta(k,n)$$

Now , from (A.1):

$$\theta(d,n+1) = \theta(d-1,n) + \theta(d-1,n+1) - \theta(d/2-1,n)$$

II- Let suppose that $u(d,n) = 2^n - 1$. Hence $2^n \le d$. Beside, from the Remark 3 that follows Lemma 2.3 we have: $l(d,n) = [d/2]$. We analyse two cases:

II.1- $2^n = d$; then $u(d,n) = d-1$ and we are in case I.

II.2- $2^n < d$; then: $n \le [\log_2 (d-1)]$ and therefore: $\theta(d-1,n) = 0$.

Beside, because $2^n < d$ , $l(d-1,n) = [(d-1)/2]$ we have:

$$\theta(d-1,n+1) = \sum_{k=[(d-1)/2]}^{2^n-1} \theta(k,n) = \begin{cases} \theta(d-1,n+1) & d \quad is \quad odd \\ \theta(d-1,n+1) - \theta(d/2-1,n) & d \quad is \quad even \end{cases}$$

We obtain the above condition from (A.4). This completes the proof.

From the above Lemma A.1 we have: $\theta(d,n+1) \le \theta(d-1,n) + \theta(d-1,n+1)$. In order to obtain an upper bound for $\theta(d,n)$, we will study the following sequence defined recursively:

$$\omega(d,n) = \omega(d-1,n-1) + \omega(d-1,n) \qquad \text{(A.5)}$$

with the following boundary conditions:

C1- $\omega(d,2) = 0 \quad \forall d \ge 4$

C2- $\omega(d,d-1) = d-2 \quad \forall d \ge 3$

It is not difficult to see that: $\theta(d,n) \le \omega(d,n)$. From now on we obtain some properties of $\omega(d,n)$.

Let consider the vector $[\omega(d,d-1), \omega(d,d-2), \ldots, \omega(d,3)]^t$. From (A.5) and conditions C1 and C2 we have:

$$\begin{bmatrix} \omega(d,d-1) \\ \vdots \\ \omega(d,3) \end{bmatrix} = \begin{bmatrix} d-2 \\ \vdots \\ 0 \end{bmatrix} + \begin{bmatrix} B_{11} & B_{12} \\ B_{21} & B_{22} \end{bmatrix} \begin{bmatrix} 0 \\ \omega(d-1,d-2) \\ \vdots \\ \omega(d-1,3) \end{bmatrix}$$

where:

$$B_{11} = [0]_{1x1}$$

$$B_{12} = \begin{bmatrix} 0 & \cdots & 0 \end{bmatrix}_{1x(d-4)}$$

$$B_{21} = \begin{bmatrix} 0 & \cdots & 0 \end{bmatrix}^{t}_{1x(d-4)}$$

$$B_{22} = \begin{bmatrix} 1 & 1 & & 0 \\ & \ddots & \ddots & \\ & & \ddots & 1 \\ 0 & & & 1 \end{bmatrix}_{(d-4)x(d-4)}$$

**Lemma A.2**: Let $d \geq 5$. Then we have:

$$\begin{bmatrix} \omega(d,d-1) \\ \omega(d,d-2) \\ \vdots \\ \omega(d,3) \end{bmatrix} = b_{d-4} + \sum_{k=0}^{d-5} A_{d-4} A_{d-5} \dots A_{d-k-4} b_{d-k-5} \qquad \text{(A.6)}$$

where:

$$A_{d-r} = \begin{bmatrix} B_{11} & B_{12} \\ B_{21} & B_{22} \end{bmatrix}$$

$$B_{11} = [0]_{(r-3)x(r-3)}\,;\; B_{21}^{t} = B_{12} = \begin{bmatrix} 0 & \cdots & 0 \end{bmatrix}_{(r-3)x(d-r)}$$

$$B_{22} = \begin{bmatrix} 1 & 1 & & 0 \\ & \ddots & \ddots & \\ & & \ddots & 1 \\ 0 & & & 1 \end{bmatrix}_{(d-r)x(d-r)}$$

$$b_{d-r} = \begin{bmatrix} 0 & \cdots & d-r+2 & \cdots & 0 \end{bmatrix}^t_{1x(d-r)}$$

where the nonzero element of $b_{d-r}$ is in position $r-3$.

**Proof**: It will be by induction on $d$.

For $d = 5$ we have:

$$\begin{bmatrix} \omega(5,4) \\ \omega(5,3) \end{bmatrix} = \begin{bmatrix} 3 \\ 0 \end{bmatrix} + \begin{bmatrix} 0 & 0 \\ 0 & 1 \end{bmatrix} \begin{bmatrix} 0 \\ \omega(4,3) \end{bmatrix}$$

Let suppose that it is true for $d$ and prove that it is also true for $d+1$:

$$\begin{bmatrix} \omega(d+1,d) \\ \omega(d+1,d-1) \\ \vdots \\ \vdots \\ \omega(d+1,3) \end{bmatrix} = \begin{bmatrix} d-1 \\ 0 \\ \vdots \\ \vdots \\ 0 \end{bmatrix} + \begin{bmatrix} 0 & 0 & 0 & \cdots & 0 \\ 0 & 1 & 1 & \cdots & 0 \\ 0 & 0 & 1 & \cdots & 0 \\ \vdots & \vdots & \vdots & \ddots & \vdots \\ 0 & 0 & 0 & \cdots & 1 \end{bmatrix} \begin{bmatrix} 0 \\ \omega(d,d-1) \\ \omega(d,d-2) \\ \vdots \\ \omega(d,3) \end{bmatrix}$$

From the hypothesis of induction, the last $d-3$ components of the vector $\begin{bmatrix} 0 & \omega(d,d-1) & \cdots & \omega(d,3) \end{bmatrix}^t$ can be written, using (A.6) as:

$$\begin{bmatrix} \omega(d+1,d) \\ \vdots \\ \omega(d+1,3) \end{bmatrix} = \begin{bmatrix} d-1 \\ \vdots \\ 0 \end{bmatrix} + A_{d-3} \left\{ b_{d-4} + \sum_{k=0}^{d-5} A_{d-4} \cdots A_{d-k-4} b_{d-k-5} \right\}$$

We have added to the element of $b_{d-4} + \sum_{k=0}^{d-5} A_{d-4} \cdots A_{d-k-4} b_{d-k-5}$ a row and/or a column of zeros to obtain the same dimension. This completes the proof.

**Lemma A.3**: Under the same conditions of Lemma A.2 let $a_{ij}$, $1 \le i, j \le d-3$, be the coefficients of matrix $A_{d-4} \cdots A_{d-r}$, then:

$$a_{ij} = \begin{cases} \binom{r-4}{i-2} & 2 \le i \le r-2 \quad and \quad j = r-2 \\ \binom{r-3}{r-3-j+i} & r-1 \le j \le d-3 \quad and \quad j-r+3 \le i \le j \\ 0 & in \quad the \quad other \quad cases \end{cases} \qquad (A.7)$$

**Proof**: It will be by induction on $r$. The property is obviously true for $r = 4$. Let suppose that it is true for $r$ and prove that it is also true for $r+1$. Let denote for $b_{ij}$ the coefficient of matrix $A_{d-(r+1)}$, then:

$$b_{ij} = \begin{cases} 1 & i = j \quad ; \quad j = r-1,\ldots,d-3 \\ 1 & i = j-1 \quad ; \quad j = r-1,\ldots,d-3 \\ 0 & in \quad the \quad other \quad cases \end{cases}$$

Let $c_{ij}$ be the coefficients of matrix $A_{d-4} \cdots A_{d-r} A_{d-(r+1)}$. Then , if $j = r-1$:

$$c_{i(r-1)} = \sum_{k=1}^{d-3} a_{ik} b_{k(r-1)} = a_{i(r-1)} b_{(r-1)(r-1)} = \binom{r-3}{r-3-(r-1)+i} = \binom{r-3}{i-2}$$

for: $2 \le i \le r-1$. In the other cases $c_{i(r-1)} = 0$. Beside, if : $r \le j \le d-3$ then:

$$c_{ij} = \sum_{k=1}^{d-3} a_{ik} b_{kj} = a_{i(j-1)} b_{(j-1)j} + a_{ij} b_{jj} = \binom{r-3}{r-2-j+i} + \binom{r-3}{r-3-j+i} = \binom{r-2}{r-2-j+i} \text{f}$$

or $j-r+2 \le i \le j$. In the other cases $c_{ij} = 0$. This completes the proof .

**Corollary A.4**: Under the same assumptions that Lemma A.3:

$$A_{d-4}\cdots A_{d-r}b_{d-(r+1)} = \begin{bmatrix} 0 \\ (d-r+1)\binom{r-4}{0} \\ (d-r+1)\binom{r-4}{1} \\ \vdots \\ (d-r+1)\binom{r-4}{r-4} \\ \vdots \\ 0 \end{bmatrix}$$

**Proof**: The only nonzero element of vector $b_{d-(r+1)}$ is in position $r-2$ and has value $d-r+1$. From Lemma A.3, the column $r-2$ of matrix $A_{d-4}\cdots A_{d-r}$ is:

$$\begin{bmatrix} 0 & \binom{r-4}{0} & \binom{r-4}{1} & \cdots & \binom{r-4}{r-4} & \cdots & 0 \end{bmatrix}^t$$

This completes the proof.

**Proposition A.5**: Let $d \geq 5$. Then:

$$\omega(d,d-i) = \sum_{n=i-2}^{d-5}(d-n-3)\binom{n}{i-2} \quad \text{for} \quad i \geq 2 \qquad \text{(A.8)}$$

**Proof**: From Lemma A.2 and Corollary A.4 we have:

$$\begin{bmatrix} \omega(d,d-1) \\ \omega(d,d-2) \\ \vdots \\ \vdots \\ \vdots \\ \vdots \\ \omega(d,3) \end{bmatrix} = \begin{bmatrix} d-2 \\ 0 \\ \vdots \\ \vdots \\ \vdots \\ \vdots \\ 0 \end{bmatrix} + \sum_{k=0}^{d-5}(d-k-3)\begin{bmatrix} 0 \\ \binom{k}{0} \\ \vdots \\ \vdots \\ \binom{k}{k} \\ \vdots \\ 0 \end{bmatrix}$$

From the above expression we obtain immediately the result.

From (A.8 ) it can be proved that for $k \geq 3$ :

$$\omega(d,k) = \frac{(2d-k-3)}{k-3}\binom{d-4}{d-k}$$

And from the above result and the definition of $\omega(d,k)$ we have:

$$\theta(d,k) \leq \frac{(2d-k-3)}{(k-3)}\binom{d-4}{d-k} \qquad \text{(A.9)}$$

## 9 References.